# Negative Index of Refraction in Optical Metamaterials


V. M. Shalaev, W. Cai, U. Chettiar, H.-K. Yuan, A. K. Sarychev, V. P. Drachev, and A. V. Kildishev

School of Electrical and Computer Engineering, Purdue University, West Lafayette, IN 47907



An array of pairs of parallel gold nanorods is shown to have a negative refractive index $n' \approx -0.3$ at the optical communication wavelength of 1.5 µm. This effect results from the plasmon resonance in the pairs of nanorods for both the electric and magnetic components of light. The refractive index is retrieved from the direct phase and amplitude measurements for transmission and reflection, which are all in excellent agreement with our finite difference time domain simulations. The refraction critically depends on the phase of the transmitted wave, which emphasizes the importance of phase measurements in finding $n'$.




One of the most fundamental notions in optics is that of refractive index, which gives the factor by which the phase velocity of light is decreased in a material compared to vacuum conditions. Negative-index materials (NIMs) have a negative refractive index so that electromagnetic waves in such media propagate in a direction opposite to the flow of energy, which is indeed unusual and counterintuitive. There are no known naturally-occurring NIMs. However, artificially designed materials (metamaterials) can act as NIMs. Metamaterials can open new avenues to achieving unprecedented physical



properties and functionality unattainable with naturally-existing materials. Proof-of-principle experiments [1] have shown that metamaterials can act as NIMs at *microwave* wavelengths. NIMs drew a large amount of attention after Pendry predicted that NIMs with a refractive index $n' = -1$ can act, at least in principle, as a 'perfect lens' allowing imaging resolution which is limited not by the wavelength but rather by material quality [2]. Materials with structural units much smaller than the wavelength can be characterized by a dielectric permittivity $\varepsilon = \varepsilon' + i\varepsilon''$ and a magnetic permeability $\mu = \mu' + i\mu''$. In this case, the real part ($n'$) of the complex refractive index ($n = n' + in'' = \sqrt{\varepsilon\mu}$) becomes negative provided that the condition [3] $\varepsilon''\mu' + \mu''\varepsilon' < 0$ is fulfilled. The condition $\varepsilon' < 0$ and $\mu' < 0$ is sufficient (but not necessary) for negative refraction [4].

While negative permittivity in the optical range is easy to attain for metals, there is no magnetic response for naturally-occurring materials at such high frequencies. Recent experiments employing lithographically-fabricated structures showed that a magnetic response and negative permeability can be accomplished in the terahertz [5 6 7] spectral ranges. These experiments showed the feasibility of optical NIMs because a magnetic response is a precursor for negative refraction. Still, the ultimate goal of negative refraction was not accomplished in these experiments. Below we report our experimental observation of a negative refractive index for the optical range, specifically, for the wavelengths close to 1.5 µm (200 THz frequency). The NIM structural design we used follows our recent theoretical prediction of negative refraction in a layer of pairs of parallel metal nanorods [8].



For a transparent optical material of thickness $\Delta$, the phase shift $\phi_t$ of the transmitted wave is equal to $2\pi n' \Delta / \lambda$ so that $n' < 0$ results in $\phi_t < 0$. In experiments using interferometry, the phase shift in a material can be precisely measured relative to a layer of air of the same thickness: $\phi_{t,\exp} = \phi_t - \phi_0$, where the phase shift in air is $\phi_0 = 2\pi\Delta/\lambda$. Then $n'$ is negative in the material provided that $\phi_{t,\exp} < -\phi_0$. In general, for a metal-dielectric material that has absorption, the relation between $\phi_t$ and $n$ is more complicated and phase measurements should be accompanied by measurements of the transmittance and reflectance amplitudes. Below we compare our experimental data on phase and amplitude in transmittance and reflectance with finite difference time domain (FDTD) simulations and retrieve the refractive index from the exact formula for $n$. Surprisingly, the simple criterion above works rather well for our samples and the retrieved refractive index is negative for $\phi_{t,\exp} < -\phi_0$.

Figure 1a schematically shows the array of closely-spaced pairs of parallel metal nanorods used in our experiments. The two parallel rods form an open current loop, which acts as a transmission line and has a current resonance. Such a loop is 'closed' through the displacement current (the blue arrows in Fig. 1a), and the structure therefore supports the resonant modes of both the electric and magnetic components of light. For normally-incident light with the electric field polarized along the rods and the magnetic field perpendicular to the pair, the electric and magnetic responses both can experience a resonant behavior at certain frequencies that depend on the rod dimensions and their separations. In this design, the electric component of the incident wave excites a symmetric current mode in the two metal rods, whereas the magnetic field component excites an anti-symmetric mode (see Fig. 1a). The excitation of such plasmon



resonances for both the electric and magnetic field components of the incident light results in the resonant response of the refractive index, which can become negative above the resonance as previously predicted [8]. This resonance can be thought of as a resonance in an optical LC-circuit, where the metal rods act as inductive elements L and the dielectric gaps between the rods act as capacitive elements C.

Experimental results were obtained using a 2 mm × 2 mm array of nanorods. Figure 1b shows a field-emission scanning electron microscope (FE-SEM) image of a portion of the sample as well as a closer view of a single pair of rods. A schematic for the whole sample is given in Fig. 1c. The first sample (Sample A) was fabricated on a 180-nm layer of indium tin oxide (ITO) coated onto glass. Another sample of the same structure (Sample B) was fabricated without the ITO layer. A JEOL JBX-6000FS electron beam writer with optimized dosage was used for writing. After the photo-resist was developed, the desired sandwich structure of metal rods and silicon dioxide ($SiO_2$) spacer was deposited in an electron beam evaporator at high vacuum. A lift-off process was then performed to obtain the array of paired nanorods.

The optical characteristics of the periodic array of nanorods shown in Fig. 1a were simulated using a 3D FDTD formulation, where the rectangular elementary cell of Fig. 1c represented the properties of the entire sample.

The core FDTD techniques have all been published extensively [9]. To simulate the permittivity, the matching Debye models were obtained from the Drude models for the



gold nanorods, $\varepsilon_{\mathrm{Au}} = 9.0 - (1.37 \times 10^{16})^2 / [\omega^2 + i(1.003 \times 10^{14})\omega]$) and ITO,

$\varepsilon_{\mathrm{ITO}} = 3.46 - (2.91 \times 10^{15})^2 / [\omega^2 + i(1.503 \times 10^{14})\omega]$ , where $\omega$ is the frequency of

incident light. These formulas provide excellent agreement with the measured optical

constants for both materials. We set an elementary cell with dimensions

$1.9 \ \mu\mathrm{m} \times 0.7 \ \mu\mathrm{m} \times 4 \ \mu\mathrm{m}$ ($x \times y \times z$) illuminated by a monochromatic plane wave at

normal incidence. The overall geometry of the nanorods inside the elementary cell

followed the trapezoidal shape of the experimental sample. Since we used a uniform

grid with a spatial resolution of 10 nm, the 5-nm Ti layers (Fig. 1c) were not taken into

account in those simulations (a higher resolution would be beyond reasonable

simulation time even for our 3D FDTD parallel code deployed at the high-end computer

cluster). The thick glass layer was considered infinite in the simulations. Two perfectly

matching layers [9], which emulate infinite propagation of the scattered field, were

arranged in front of and behind the nanorods. Standard periodic boundary conditions

were applied to the elementary cell elsewhere to ensure periodicity of the entire array.

The total-scattered field separation technique was used to obtain the scattered fields [9].

The field data for reflection and transmission coefficients were taken at selected

evaluation planes in front of and behind the nanostructure using one-period sampling

and additional Fourier filtering with averaging over the evaluation planes.

The complex values of the reflected and transmitted electric fields, $E_r$ and $E_t$, are

calculated using 3D FDTD for normally-incident light with electric field $E_i$. Then,

$r = \alpha \, E_r \left(-d\right) \big/ E_i \left(-d\right)$ and $t = \alpha \, E_t \left(d\right) \big/ E_i \left(-d\right)$ are obtained from the field values,

where $\alpha = \exp\left[ik\left(\Delta - 2d\right)\right]$, $k$ is the wavenumber in air and $d$ is the distance from

the center of the layer of the paired nanorods, with the total thickness $\Delta = 160 \ \mathrm{nm}$ , to



the field evaluation planes in front and behind the sample. The distance $d$ is chosen so that the reflected and transmitted waves ($E_r$ and $E_t$) are plane waves with no more than 1% deviation in magnitudes in the evaluation planes.

For a single layer, the complex index of refraction ($n$) of the nanorod layer can be found from [10]:

$$\cos nk\Delta = \frac{1 - r^2 + t^2}{2t}. \tag{1}$$

Eq. (1) provides the solution for retrieving the refractive index of a given thin layer of a passive material ($n'' > 0$ and $\mathrm{Re}(Z) > 0$, where $Z$ is the impedance) with unknown parameters [10]. Specifically, we obtain the impedance (not shown here) and the refractive index of the equivalent homogeneous layer with the same complex reflectance $r$ and transmittance $t$ as the actual array of nanorods. Such a homogeneous layer with equivalent $n$ and $Z$ gives the same far-field distribution outside the sample as the actual layer of metamaterial. Note that for a thin layer such retrieval can be performed unambiguously.

Figure 2 shows results of our experimental measurements of transmittance (top), and reflectance (bottom) for light propagating through the sample with ITO layer (Sample A). The transmission ($T = |t|^2$) and reflection ($R = |r|^2$) spectra are measured with a Lambda 950 spectrophotometer from Perkin-Elmer using linearly polarized light. Two different light polarizations have been used, one with the electric field parallel the



nanorod pair major axis and the other with the electric field perpendicular to this axis. The transmission spectra are collected at normal incidence, and the reflection spectra are measured at a small incident angle of 8º. The transmittance for Sample A is normalized by the transmittance of the ITO-glass substrate for both light polarizations. The refractive index of the glass is 1.48.

As seen in Fig. 2, the system of parallel gold nanorods shows a strong plasmonic resonance near 1.3 μm for light with the electric field polarized parallel to the rods. In this case, light excites both electric and magnetic responses. For light with the electrical field polarized perpendicular to the rods, only the electric (transverse) plasmon resonance is excited, with a resonant wavelength near 800 nm, and the magnetic response is negligible in this case. We note good qualitative agreement between our 3D FDTD simulations and experimental data – both the positions and the widths of the resonances are reproduced well. For simulations of metals in the optical range, which are known to be a hard problem, the accomplished level of agreement is very good. To investigate the refraction of our material, we also performed phase measurements using polarization and walk-off interferometers. These interferometry methods are capable of revealing that the phase and group velocities of light are antiparallel in NIMs, rather than being parallel as in normal, positive-index materials. In the polarization interferometer, two optical channels have a common geometrical path and differ by the polarization of light. This allows one to measure the phase difference between orthogonally-polarized waves $\Delta\phi = \phi_\parallel - \phi_\perp$ caused by anisotropy of a refractive material. Note that the phase acquired in the substrate does not contribute to the phase difference $\Delta\phi$. The walk-off interferometer has two optical channels which differ in



geometrical paths; it gives a phase shift introduced by a sample ($\phi_s$) relative to a reference ($\phi_r$): $\delta\phi = \phi_s - \phi_r$. A layer of air with the same thickness as the layer of rods was used as the reference, i.e., $\phi_r = \phi_0$. Both the reference and the sample beams go through the substrate so that the phase acquired in the substrate does not contribute to the measured phase shift $\delta\phi$ if the substrate has no variations in optical thickness. The walk-off effect in calcite crystals is employed to separate the two beams and then bring them together to produce interference. The phase shifts $\delta\phi_\parallel$ and $\delta\phi_\perp$ are measured for the two light polarizations, using a set of diode lasers and a tunable erbium laser, and their difference is compared with the phase anisotropy $\Delta\phi$ obtained from polarization interferometry (note that $\Delta\phi = \delta\phi_\parallel - \delta\phi_\perp$). The instrumental error of the phase anisotropy measurement by polarization interferometer is ±1.7º. We note that variations in the substrate thickness do not affect the results of our phase anisotropy measurements, which is typical for common path interferometers. In the case of the walk-off interferometer, the thickness variation gives an additional source of error, causing the error for the absolute phase shift measurements to increase up to ±4º. The high accuracy of phase measurements (about $0.005\lambda$, for the phase anisotropy, and $0.01\lambda$, for the absolute phase shift) is common for interferometry.

Figure 3 shows results of the phase measurements, which are compared with our simulations. We note good agreement between our FDTD simulations and experimental data; the phase shifts at two different wavelength segments are reproduced very well. As seen in Fig. 3, in the spectral range between 1 and 1.6 µm the absolute phase shift $\delta\phi_\perp$ measured with respect to the layer of air is positive and nearly flat (close to +10º) in



magnitude. In contrast, $\delta\phi_{\parallel}$ shows a strong resonant dependence such that the difference $\delta\phi_{\parallel} - \delta\phi_{\perp}$ agrees very well with our measurements of $\Delta\phi$ using polarization interferometry. According to Fig. 3, the phase difference $\Delta\phi = \phi_{\parallel} - \phi_{\perp} = \delta\phi_{\parallel} - \delta\phi_{\perp} \approx -50°$ and the phase shift $\delta\phi_{\parallel} = \Delta\phi + \delta\phi_{\perp} \approx -40°$ at 1.2 µm. The qualitative criterion discussed above requires that the experimental phase shift be less than $-\phi_0 = -2\pi\Delta/\lambda = -49°$ at 1.2 µm, so the observed $\delta\phi_{\parallel}$ suggests that the refractive index is expected to be close to zero but positive in this case.

To retrieve the exact values of the refractive index we generalized Eq. (1) for a multi-layer system, following the approach of Ref. 10, and took into account the ITO layer (to be published elsewhere). Fig. 4a shows the obtained index and demonstrates excellent agreement between measurements and simulations. As expected, the refractive index is close to zero between 1 and 2 µm (the lowest $n'$ is 0.08 at 1.1 µm) but it remains positive. Clearly, the system of parallel gold nanorods shows a strong plasmonic resonance near 1.3 µm for light with the electric field polarized parallel to the rods. In this case, light excites both electric and magnetic responses, resulting in the anomalously low refractive index.

To understand the effect of ITO, the refractive index was retrieved from Eq. (1) for the same nanorod array but in free space, i.e., *without* the ITO-coated glass substrate. To estimate the possible error of the inversion (Eq. (1)), the standard deviation ($\sigma$) of the refractive index was found using the deviation of reflected and transmitted fields obtained at the evaluation planes. Our results show a negative $n'$ in this case. The inset



in Fig 4b gives a closer view of the wavelength domain for negative $n'$ with upper and lower boundaries of $n' \pm \sigma$. The refractive index is negative in the range between 1.1 μm and 1.4 μm, reaching a magnitude of $n' \approx -0.15$ at $\lambda = 1.25\ \mu m$. The relatively small magnitude of the negative $n'$ can be enlarged by increasing the metal filling factor [8]. The imaginary part of the refractive index also shows resonant behavior and it is large near the resonance. We believe that by optimizing the system, losses can be decreased. One of the ways to decrease losses is to fabricate structures much smaller than the wavelength such that the radiation loss would decrease.

In order to prove that a sample without an ITO layer may indeed have a negative index of refraction, we fabricated a similar array of nanorods directly on glass (for this Sample B, the sizes of the lower nanorods are slightly larger, 780 nm × 220 nm, and the sizes of the elementary cell are slightly less, 1800 nm × 640 nm; the metal filling factor is 13.5% which exceeds the 8.3% filling factor for Sample A). The slightly larger rod sizes resulted in a small shift of the resonance toward longer wavelengths, whereas the absence of the ITO layer and a larger metal filling factor facilitate the negative refraction, as seen in Fig. 5.

In Fig. 5a we show the results of our phase measurements for Sample B, which were performed with diode lasers and a tunable erbium laser. The detected phases are in excellent agreement with simulations. The inset depicts the measured transmittance and reflectance magnitudes (for the parallel polarization) verified with the erbium laser in the spectral range of interest where n is negative. Fig. 5b exhibits $n'$ obtained with the



use of Eq. 1. [ $n''$ is similar to that in Fig. 4 and it is not shown here.] The inset of Fig. 5b provides a zoomed view, where we show the quadratic least square fit for our experimental data (dashed line) and compare it to the simulated data (solid line). Our simulations (for both Samples A and B) suggest a relatively weak sensitivity of $n'$ to variations in reflectance and transmittance amplitudes and its critical dependence on phases (this will be discussed in more detail elsewhere). The excellent agreement between calculated and measured phase shifts resulted in close values for the simulated and experimental values of $n'$. We note that the obtained absolute phase shift of -61° (not shown) in the light transmittance at $\lambda = 1.5\,\mu m$ is well below the critical phase $-\phi_0 = -40^\circ$ at $1.5\,\mathrm{\mu m}$, so the phase criterion used above also predicts negative refraction in this case. The refractive index is negative between 1.3 μm and 1.6 μm, with $n' = -0.3 \pm 0.1$ at $\lambda = 1.5\,\mu m$. The spectral range of a negative $n'$ is shifted with respect to the resonance providing a rather high transmittance of about 20%.

In conclusion, for an array of pairs of parallel gold rods, we obtained a negative refractive index of $n' \approx -0.3$ at the optical communication wavelength of 1.5 μm. This new class of negative-index materials (NIMs) is relatively easy to fabricate on the nanoscale and it opens new opportunities for designing negative refraction in optics. The frequency for negative refraction depends on both the size of the metal rods and their separation. The negative-refraction frequency can span the visible and near-infrared parts of the spectrum through appropriate nanorod array design. Further optimization of the proposed structures would allow NIMs with lower losses and larger



magnitudes of negative refraction, resulting in new applications based on this unique phenomenon.

This work was supported in part by NSF-NIRT award ECS-0210445 and by ARO grant W911NF-04-1-0350.

––––––––––––––––––––

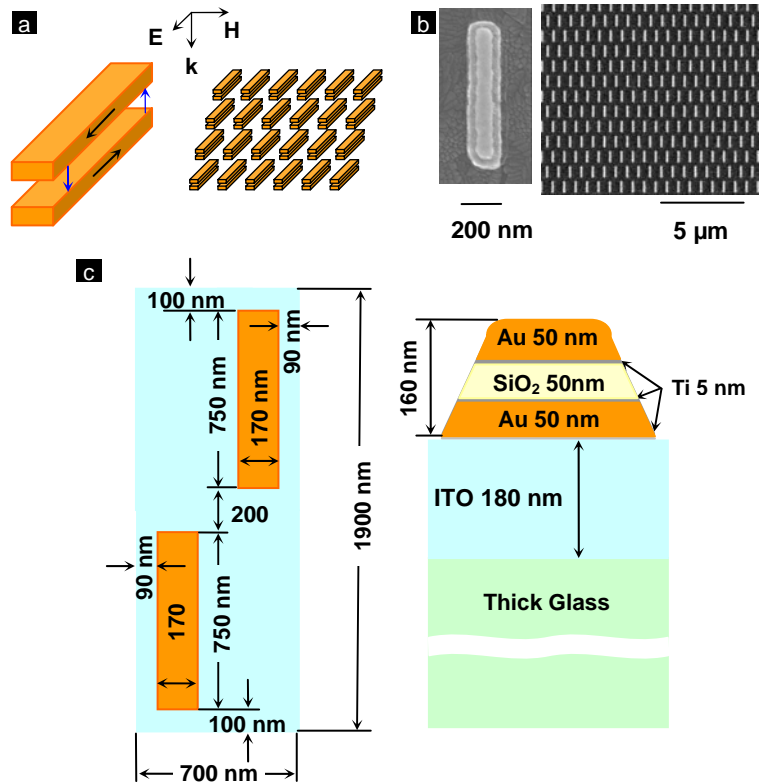

FIG. 1. Pairs of parallel gold nanorods. (a) A schematic for the array of nanorods; the H-field induced current in the rods (black arrows) is closed by displacement current (blue arrows). (b) Field-emission SEM picture (top view) of a single pair of nanorods (left) and a fragment of 2 mm × 2 mm array of pairs of nanorods (right). (c) Sizes of the lower nanorods and their separations in the array (left) and a side-view schematic for one pair of nanorods on a substrate formed by an ITO layer and glass (right).



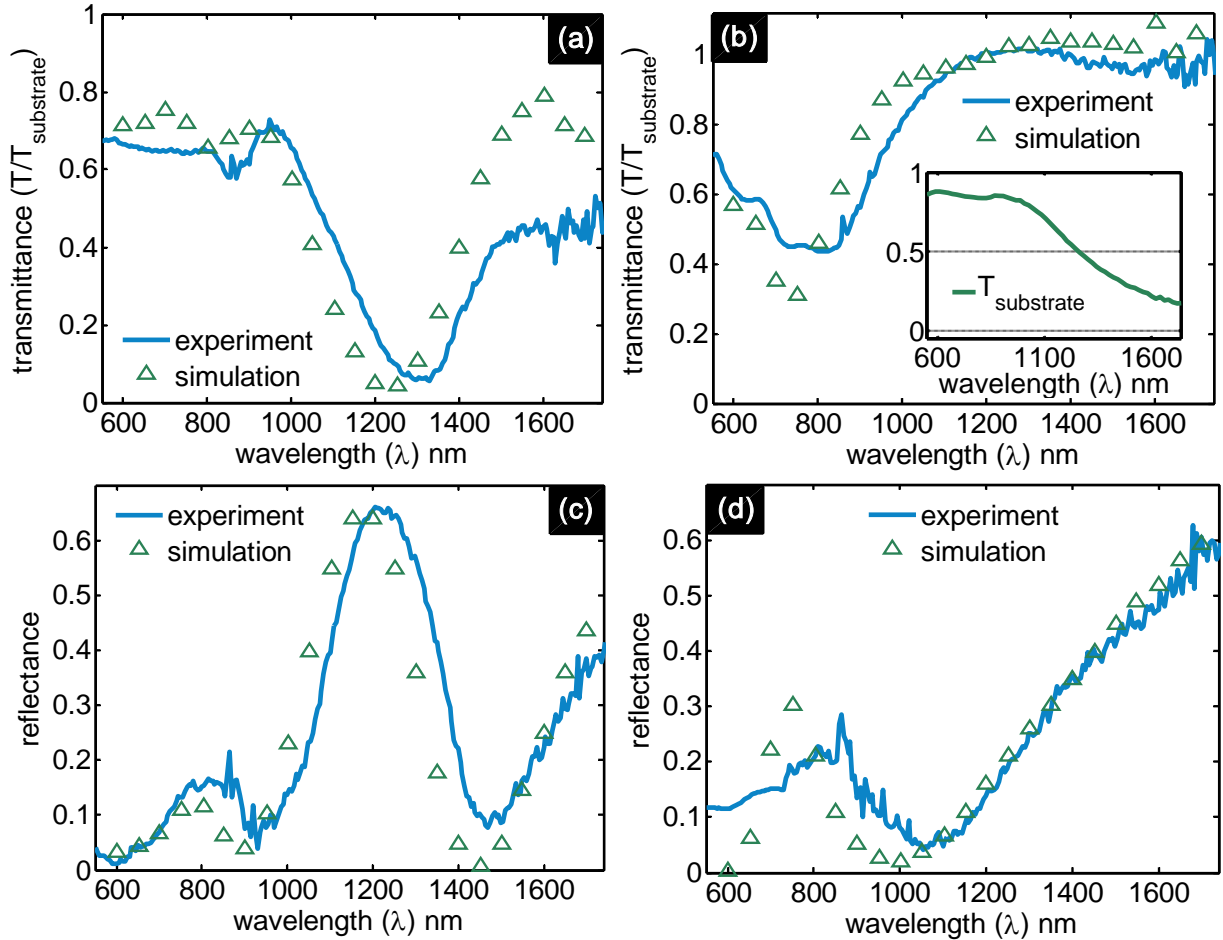

**FIG. 2. Normalized transmittance T/T$_{substrate}$ ((a) and (b)) and reflectance R ((c) and (d)) of light for Sample A. Two different polarizations are used, one with the electric field along the rods, as in (a) and (c), and one perpendicular to the rods, as in (b) and (d). The transmittance of the substrate (ITO-coated glass) is shown as inset in (b).**



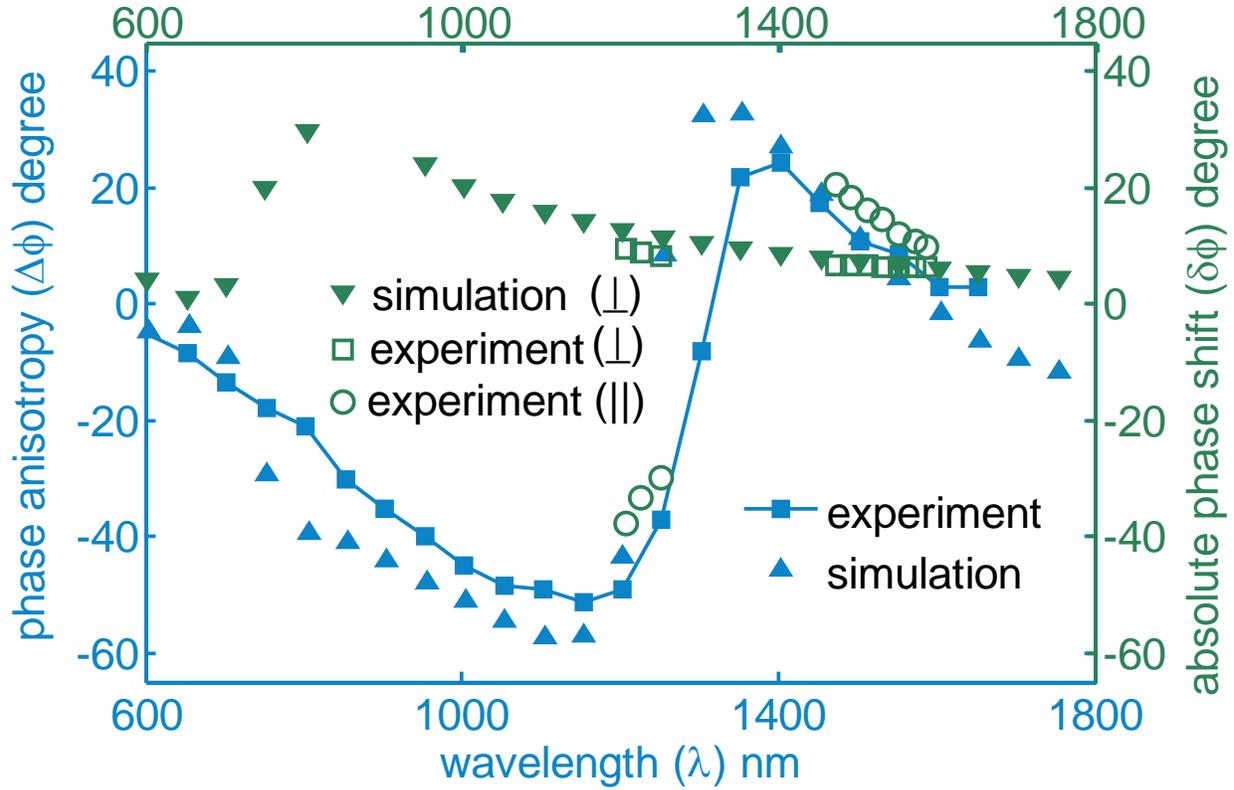

FIG. 3. Phase change in transmission by the layer of pairs of gold rods of Sample A, for light polarized parallel and perpendicular to the rods. The measured (■) and simulated (▲) phase anisotropies $\Delta\phi = \phi_\parallel - \phi_\perp = \delta\phi_\parallel - \delta\phi_\perp$ are shown (left ordinate). The absolute phase shifts $\delta\phi_\parallel$ and $\delta\phi_\perp$ for waves with parallel (○) and perpendicular (□) polarizations are also depicted (right ordinate). The absolute shifts $\delta\phi_\perp$ for the perpendicular polarization are compared to simulation results (▼); $\delta\phi_\parallel$ can be found as $\delta\phi_\parallel = \Delta\phi + \delta\phi_\perp$.



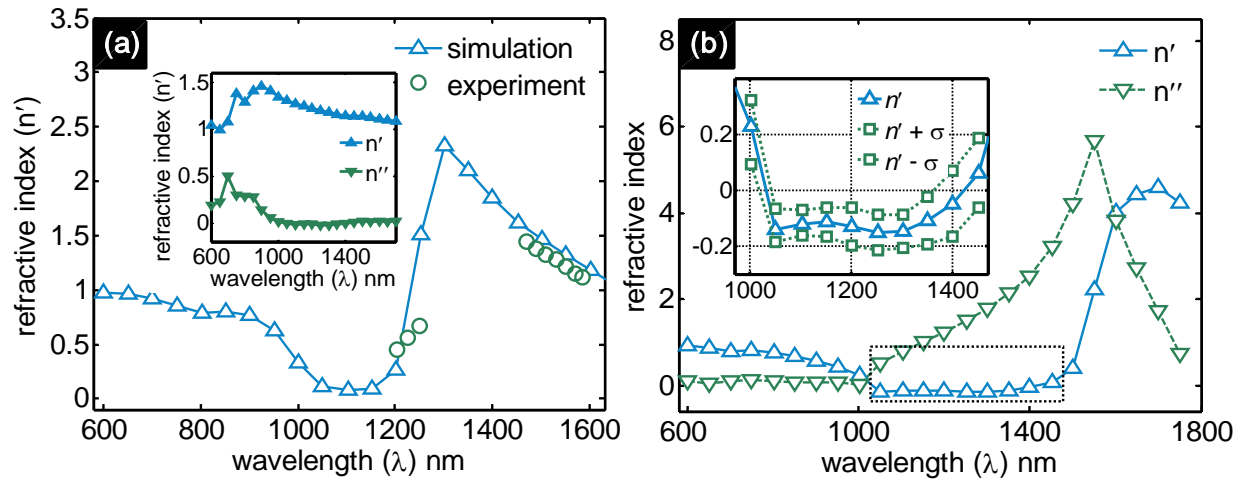

**FIG. 4. (a) The real part of refractive index for the parallel polarization of incident light restored from measurements and simulations for Sample A. The inset in (a) shows the refractive index calculated for the perpendicular polarization. (b) Refractive index for Sample A simulated without ITO-glass substrate. The inset in (b) shows the zoomed view of Fig. 4b for $n' < 0$, including the upper and lower bounds with a one-sigma deviation.**



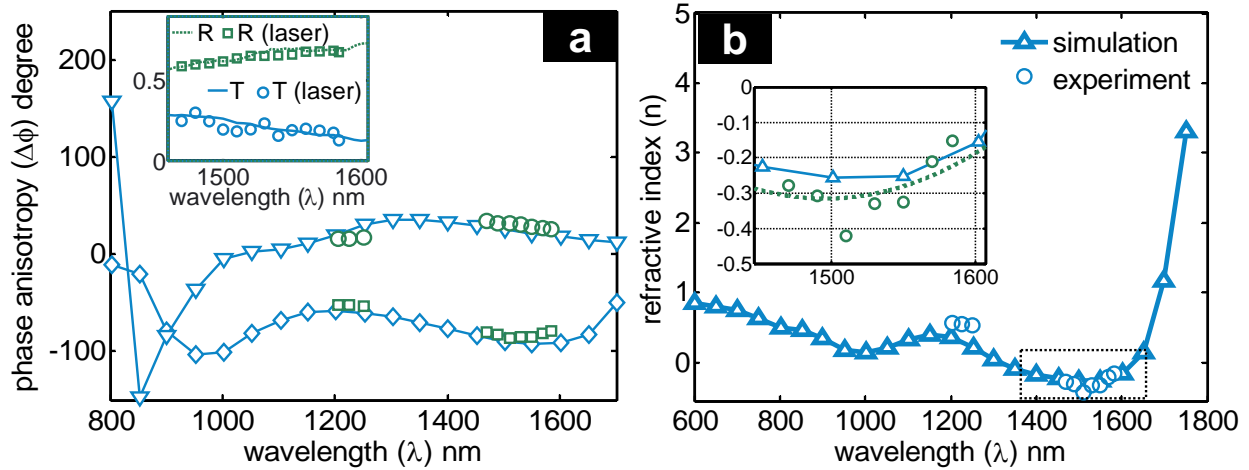

**FIG. 5.** (a) Phase anisotropy $\Delta\phi = \phi_\parallel - \phi_\perp$ for transmitted ($\diamond$ simulation and $\square$ experiment) and reflected ($\nabla$ simulation and $\bigcirc$ experiment) light. The inset in (a) shows measured transmittance and reflectance magnitudes for the parallel polarization verified with laser measurements. (b) The real part of the refractive index restored from experimental data and compared to simulations. The inset depicts a zoomed view of the region of negative refraction; the dashed and solid lines show the trends for experimental data ($\bigcirc$) and simulated data ($\triangle$).